\documentclass[a4paper,11pt]{article}

\usepackage{jheppub} 

\usepackage[T1]{fontenc} 

\usepackage{graphicx} 
\usepackage{amsthm}

\newcommand{\abs}[1]{\left| #1 \right|} 
\newcommand{\ket}[1]{| #1 \rangle } 
\newcommand{\bra}[1]{\langle  #1 |} 
\let\baraccent=\= 
\renewcommand{\=}[1]{\stackrel{#1}{=}} 
\newcommand{\p}{\prime}

\def\N{\mathcal{N}}

\def\L{\mathcal{L}}
\def\H{\mathcal{H}}
\def\R{\mathcal{R}}

\def\Tr{\text{Tr}}
\def\rank{\text{rank}}

\newtheorem{thm}{Theorem}[section]

\newtheorem{defn}[thm]{Definition}

\newtheorem{Proposal}[thm]{Proposal}

\title{Probing the Page transition via approximate quantum error correction}

\author[a,b]{Haocheng Zhong}

\affiliation[a]{Shing-Tung Yau Center of Southeast University, Nanjing 210096, China}
\affiliation[b]{School of Physics, Southeast University, Nanjing 211189, China}

\emailAdd{zhonghaocheng@outlook.com}

\abstract{In recent years, there is a huge progress in understanding the black hole information problem, and the key is that the black hole entropy of radiation should be calculated by the island formula, which describes the Page curve to ensure the unitarity of black hole evaporation. In the paper, we find that the black hole evaporation interpreted by the island formula can be understood in the language of approximate quantum error correction. Furthermore, the Page transition, as a special property of the Page curve, should be understood as the property of approximate quantum error correction itself, i.e. a general class of quantum systems under certain conditions from approximate quantum error correction can also exhibit phenomenon similar to the Page transition.}

\begin{document} 
\maketitle
\flushbottom

\section{Introduction}

The black hole information paradox \cite{Mathur:2009hf,Almheiri:2012rt,Penington:2019npb,Almheiri:2020cfm} refers to the conflict between two fundamental principles of physics: general relativity and quantum mechanics. The paradox gained significant attention when Stephen Hawking proposed in the 1970s that black holes could emit radiation now known as Hawking radiation which would imply that black holes slowly lose mass and eventually evaporate entirely \cite{Hawking:1975vcx,Hawking:1976ra}. However, the radiation emitted by black holes appears to be random and independent of the information initially contained within them, which raises questions about the fate of the information that fell into the black hole. On the other hand, quantum mechanics states that information cannot be destroyed, and it must be conserved over time, which are the requirements of \emph{unitarity} of quantum mechanics. This implies that information about an object that falls into a black hole should somehow be preserved rather than lost. This contradiction between the predictions of general relativity and quantum mechanics is what constitutes the black hole information paradox.

The key to the paradox is that unitarity requires the fine-grained entropy\footnote{Here we are referring to the von Neumann entropy, which is a valid measure of entanglement between systems, but it is only valid when the total system is \emph{pure}. When the total system is \emph{mixed}, many quantities are proposed to replace the role of von Neumann entropy to measure entanglement, e.g. the reflected entropy \cite{Dutta:2019gen}, the balanced partial entanglement \cite{Wen:2021qgx,Camargo:2022mme,Wen:2022jxr} from partial entanglement entropy \cite{Lin:2023rxc,Lin:2024dho,Wen:2020ech,Wen:2019iyq,Han:2019scu,Wen:2018mev,Wen:2018whg}.} of radiation to satisfy the Page curve \cite{Page:1993wv,Page:2013dx} if a black hole is formed by a pure state, contradicting Hawking's calculation that the entropy of radiation should increase monotonically. The entropy of radiation $S(\rho_{ R})$, where $\rho_{ R}$ is the density matrix of radiation, should increase to reach a maximum value and then decrease back to zero. At early time of black hole, radiation starts such that the entropy increases. However, purity of the initial state should be preserved under unitary process which ensures that the entropy should become zero when the black hole finally evaporates. The argument involves a critical point called the \emph{Page time} denoted by $t_{Page}$ at which stage the trend of the entropy of radiation turns, and we call the transition as the \emph{Page transition}.

\begin{figure}[ht]
	\centering
	\includegraphics[scale=0.4]{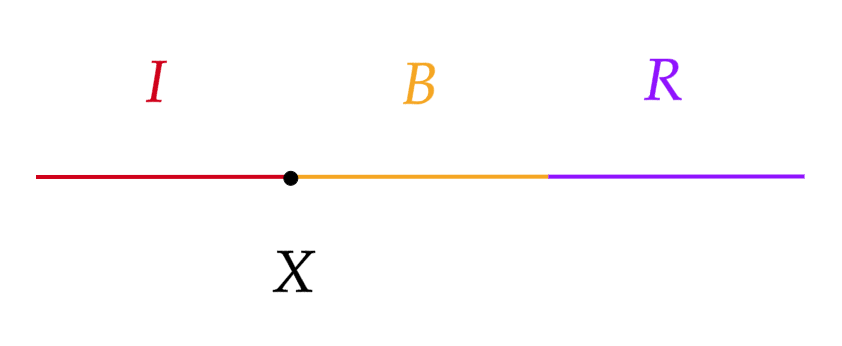}
	\caption{At a fixed time slice, the spatial region (a Cauchy slice) containing a black hole is divided into three subregions: $I,B,R$ with $X$ being the entangling surface between $I$ and $B$. The left endpoint of $I$ refers to the singularity of the black hole and the right endpoint of $R$ refers to the spatial infinity.}
	\label{fig:island}
\end{figure}

In recent years, huge progress \cite{Penington:2019npb,Almheiri:2019hni,Penington:2019kki,Almheiri:2019qdq,Almheiri:2019psf} has been made which claims the entropy of radiation should be calculated by the \emph{island formula}. To be specific, we consider a configuration at a fixed time slice as illustrated in Fig \ref{fig:island}. The spatial region is divided into three subregions: $I,B,R$ where $I$ denotes the \emph{island}\footnote{For island defined in other theories, one can see for example \cite{Takayanagi:2011zk,Fujita:2011fp,Suzuki:2022xwv,Sully:2020pza} for holographic BCFT, \cite{Chandra:2024bkn,Lin:2023ajt,Basu:2023wmv,Wen:2024uwr} for holographic Weyl transformed CFT, and see \cite{Deng:2022yll,Geng:2022slq,Geng:2024xpj,Miao:2022kve,Geng:2020qvw,Geng:2023qwm,Omidi:2021opl,RoyChowdhury:2022awr,RoyChowdhury:2023eol} for an incomplete list of recent developments.} which is located near the black hole, and $R$ represents a non-gravitational region\footnote{or in the semi-classical description where we can consider \emph{effective field theory} in the curved spacetime.} where observers within collects the radiation from the black hole horizon, and $B$ is compliment of the other two. The entropy of radiation should then be calculated by
\begin{align}\label{islandFormula}
	S(\rho_{ R})=\text{min}\left\{\text{ext}_{X}\left[\frac{Area(X)}{4G}+S_{bulk}(\tilde{\rho}_{I\cup  R})\right]\right\}
\end{align}
where $\rho_R$ denotes the reduced density matrix of $R$ defined by tracing out degrees of freedom on $I\cup B$ in the complete quantum gravity description, and $\tilde{\rho}_{I\cup  R}$ denotes the reduced density matrix of $I\cup\R$ defined by tracing out degrees of freedom on $B$ in the semi-classical description, and $X$ denotes the entangling surface between $I$ and $B$. The condition $"\text{ext}_{X}"$ says that $X$ should be chosen such that the sum of two terms on the right-hand side should be extremized under infinitesimal configuration perturbation of $X$, in which case $X$ is called the \emph{quantum extremal surface} (QES). The condition ``min'' says that if there exists multiple QES's, we choose the one which minimizes the sum of terms on the right-hand side.

Does the island formula \eqref{islandFormula} satisfy the Page curve? One can argue \cite{Penington:2019npb,Penington:2019kki,Almheiri:2020cfm,Almheiri:2019qdq} that the island formula can be understood as a competition of two saddle-point solutions of the extremal condition in the gravitation path integral description: the vanishing island and the non-vanishing island with $X$ slightly deviated from the black hole horizon, then \eqref{islandFormula} can be equivalently formulated into
\begin{equation}\label{islandFormula2}
	S(\rho_R)=\min\{ S^{island}_R,S^{no-island}_R\}
\end{equation}
The two terms on the right-hand side of \eqref{islandFormula2} behave very differently as illustrated in Fig \ref{PageCurve}. $S^{no-island}_R$ increases monotonically with time, which coincides with Hawking's calculation, while $S^{island}_R$ decreases monotonically after the formation of $X$, which is slightly later than the formation of black hole,  capturing the thermodynamic properties of the black hole. These two solutions are course-grained, i.e. the entropy of radiation as the fine-grained entropy is then upper-bounded by the two solutions such that it does follow the property of the Page curve. For more discussions about black hole information paradox, one is encouraged to consult \cite{Almheiri:2020cfm}.

Several comments can be made from the above description:
\begin{itemize}
	
	\item Before $t_{Page}$, the island does \emph{not} contribute to $S(\rho_{ R})$, but the island itself does exist. To repraise, no island has been observed from the viewpoint of observers in $R$.

	\item After $t_{Page}$, the island appears from the viewpoint of observers in $R$.

	\item The Page transition can now be understood as the emergence of the island from the viewpoint of observers in $R$.

\end{itemize}

These comments are the starting points of the paper. Later we will see that motivated by the ideas above, the Page transition can be understood via the notion of \emph{approximate quantum error correction} (AQEC) from quantum information theory, and more importantly, it is the property of AQEC itself. In other words, \emph{there can be a Page transition, which exhibits the emergence of an island from the viewpoint of some local observers, for a general class of quantum systems which obey some criteria of AQEC.} For relevant applications to black hole information paradox via the language of quantum information, one can consult \cite{Hayden:2017xed,Hayden:2018khn,Chen:2019gbt,Brown:2019rox,Akers:2019lzs,Akers:2022qdl,Akers:2021fut}.

Roughly speaking, whether the island is visible for observers in $R$ is related to the notion of \emph{reconstruction} inside the island. Parallely to the information paradox, a series of work \cite{Almheiri:2014lwa,Dong:2016eik,Harlow:2016vwg,Kamal:2019skn} constitutes a theorem, which we call the \emph{reconstruction theorem}. The theorem is originally proved for explaining the problem of bulk reconstruction in AdS/CFT, see for example \cite{Harlow:2018fse} for introduction, but the proof itself is based on notions of quantum information, and later we will argue that it is applicable for the black hole reconstruction problem. What the theorem can do is to translate the reconstruction criterion into several equivalent statements, and amazingly, one of which is closely related to AQEC. Later in the paper we will directly examine that the black hole evaporation explained by the island formula \eqref{islandFormula} can be interpreted as some AQEC process. More importantly, this AQEC process is independent of whether a black hole exists hence the process describing Page transition can be applied to more generic situations. The rest of paper aims at solidating the claim more formally.

The paper is organized as follows. In section \ref{SecQI}, we introduce some notions from quantum information, namely quantum channel, (approximate) quantum error correction and the reconstruction theorem, forming the basics for the paper. In section \ref{SecMain}, which is the main part of the paper, we examine the relation between black hole evaporation and approximate quantum error correction. We end with section \ref{SecDiscussions} where we give a proposal for our generalization and discuss some related aspects. In appendix \ref{secProof} we present a proof for the reconstruction theorem for readers who are interested.

\section{Preliminaries}\label{SecQI}

\subsection{Notation}\label{SecNotation}

Before we begin, we need to clarify our notations and conventions. We use $\mathcal{H}$ to denote the Hilbert space of a system in the relevant content, and the set of linear operators acting on $\H$ is denoted as $\mathcal{L}(\H)$. The set of density matrices in $\H$ is denoted as $s(\H)\subset \mathcal{L}(\H)$. If a system is labeled by a letter, e.g. $A$, then we write $\H_A$ as the corresponding Hilbert space and $\abs{A}$ as the dimension of $\H_A$. States and operators on $\H_A$ are also labeled with a subscript, e.g. $\ket{\psi}_A\in\H_A,O_A\in\L(\H_{A})$. The product sign $\otimes$ denotes either tensor product for states, e.g. $\ket{\psi}\otimes\ket{\phi}$, or Kronecker product for operators, e.g. $O_A\otimes I_{B}$ where $I$ is the identity. For abbreviation, we will simplify them as $\ket{\psi}\ket{\phi}$ and $O_A$ respectively, but sometimes we will keep the sign explicit to remind the readers. Whether $O_A$ is an operator on $\H_A$ or a larger space should be distinguished by the states it acts on.

\subsection{Quantum channel}

We first introduce the notion of \emph{quantum channel}. The content is based on \cite{Chatwin-Davies:2021nhs} which readers are encouraged to consult for more detailed discussions.

 \begin{defn}
 	Given two systems $A,B$, a \emph{quantum channel} $\N$ from $A$ to $B$ is a linear, completely positive and trace-preserving (CPTP) map from density matrices to density matrices:
 	\begin{equation}
 		\begin{aligned}
 			\N:s(\H_A)&\rightarrow s(\H_B)	\\
 			\rho_A&\mapsto\rho_B
 		\end{aligned}
 	\end{equation}
\end{defn}
\emph{Linearity} is to make sure that an ensemble maps to another ensemble with the same probability distribution, i.e. 
\begin{equation}
	\N(p_i\rho_i)=p_i\N(\rho_i)
\end{equation}
\emph{Trace-preserving} ensures that an operator with unital trace gets mapped to another operator with unital trace, and a map between operators is \emph{positive} if it sends a positive semi-definite operator to another positive semi-definite operator. Both properties should be required because density matrices are semi-definite and have unital trace by definition. Instead of positivity, we require a stronger condition called \emph{completely positivity}:
\begin{equation}
	\N\otimes I_E: \rho_A\otimes \sigma_E\mapsto\rho_B\otimes \sigma_E,\quad \forall \sigma_E \in s(\H_E)
\end{equation}
is positive for any auxiliary system $E$. Practically speaking, completely positivity is to ensure that the channel works suitably even though the system is coupled with environment, which is usually the case in an actual computation process.

Two examples of quantum channel are useful in the paper:
	
	\begin{itemize}
		\item For any unitary $U\in\L(\H)$, there exists a quantum channel $\N_U$ from $\H$ to itself:
		\begin{equation}\label{NU}
			\begin{aligned}
				\mathcal{N}_U: s(\mathcal{H}) &\rightarrow s(\mathcal{H}) \\
				\rho &\mapsto U \rho U^{\dagger}
			\end{aligned}
		\end{equation}

		\item For any \emph{isometry} $V: \H_A\rightarrow \H_B\otimes \H_E$, there exists a quantum channel $\mathcal{N}_V$ from $A$ to $B$:
		\begin{equation}\label{NV}
			\begin{aligned}
				\mathcal{N}_V: s(\mathcal{H}_A) &\rightarrow s(\mathcal{H}_B) \\
				\rho_A &\mapsto\Tr_E\left( V \rho_A V^{\dagger}\right)
			\end{aligned}
		\end{equation}
		where ``isometry'' means $V^\dagger V=I_A$ and $VV^\dagger=\Pi_{BE}$, i.e. projection onto $V(\H_A)\subset\H_B\otimes \H_E$.

	\end{itemize}

In fact, for any quantum channel $\N$ there exists an isometry $V$ satisfying \eqref{NV}, which is also called the \emph{isometric dilation} of $\N$, if a suitable auxiliary system $E$ is chosen \cite{Chatwin-Davies:2021nhs}.

One important feature of quantum channel is that, the \emph{relative entropy} monotonically decreases under a quantum channel, which is described as follows,
\begin{thm}\label{relentropymono}
	Let $\N$ be a quantum channel from system $A$ to system $B$, then we have
	\begin{equation}
		S(\rho|\sigma)\geq S(\N(\rho)|\N(\sigma)),\quad \forall \rho,\sigma\in s(\H_A)
	\end{equation}
	where $S(\rho|\sigma)$ is the relative entropy between $\rho$ and $\sigma$ (likewise for $S(\N(\rho)|\N(\sigma))$).
\end{thm}

\subsection{Approximate quantum error correction}
	
	Our next important notion is the \emph{quantum error correction} (QEC), which is to protect computing process against errors (or noise) which may affect the outcome of computation. One basic idea is \emph{encoding} the system into a larger one such that the original information can be protected when the enlarged system is only partly affected. One is encouraged to consult \cite{nielsen2010quantum,PreskillNotes} for a general introduction.

	We now introduce one QEC example for comprehension: the three-qutrit code \cite{Almheiri_2015}, to illustrate the ideas of QEC. Suppose Alice wants to send Bob a state based on a qutrit:
	\begin{equation}
		\ket{\psi}=a_0\ket{0}+a_1\ket{1}+a_2\ket{2}=\sum_{i=0}^{2}a_i\ket{i}
	\end{equation}
	If Bob's apparatus can \emph{not} get access to \emph{one} qutrit, then since the state is based on one qutrit, he can no longer receive the information, i.e. the qutrit is erased from Bob's viewpoint. A better idea is that Alice should send a three-qutrit state instead of a single qutrit state:
	\begin{equation}
		|\tilde{\psi}\rangle=\sum_{i=0}^{2}a_i|\tilde{i}\rangle
	\end{equation}
	where we define a set of new bases,
	\begin{equation}\label{threequtrits}
		\begin{aligned}
			& |\widetilde{0}\rangle=\frac{1}{\sqrt{3}}(|000\rangle+|111\rangle+|222\rangle) \\
			& |\widetilde{1}\rangle=\frac{1}{\sqrt{3}}(|012\rangle+|120\rangle+|201\rangle) \\
			& |\widetilde{2}\rangle=\frac{1}{\sqrt{3}}(|021\rangle+|102\rangle+|210\rangle) .
		\end{aligned}
	\end{equation}
	which spans the three-dimensional subspace of the total three-qutrit space $\H$, and we call it the \emph{code subspace} denoted by $\H_{code}\subset\H$. Nows if Alice sends this alternative state $\ket{\tilde{\psi}}$ to Bob, and we assume that Bob still can not get access to one of the three qutrits. Without loss of generality, we assume that it is the third qutrit which is erased. Bob can access to the first two qutrits, and he should implement the following operation $U_{12}$ to the first two qutrits:
	\begin{equation}
		\begin{array}{ccc}
			|00\rangle \mapsto|00\rangle & |11\rangle \mapsto|01\rangle & |22\rangle \mapsto|02\rangle \\
			|01\rangle \mapsto|12\rangle & |12\rangle \mapsto|10\rangle & |20\rangle \mapsto|11\rangle \\
			|02\rangle \mapsto|21\rangle & |10\rangle \mapsto|22\rangle & |21\rangle \mapsto|20\rangle
		\end{array}
	\end{equation}
	which ``transforms'' the new basis into the original basis as follows,
	\begin{equation}\label{trans3q}
		\left(U_{12} \otimes I_3\right)|\widetilde{i}\rangle=|i\rangle \otimes\ket{\chi} 
	\end{equation}
	where $\ket{\chi}=\frac{1}{\sqrt{3}}(|00\rangle+|11\rangle+|22\rangle)$. Therefore, when Bob applies $U_{12}$ to $|\tilde{\psi}\rangle$, he finds that
	\begin{equation}\label{trans3q2}
		\left(U_{12} \otimes I_3\right)|\widetilde{\psi}\rangle=|\psi\rangle \otimes \ket{\chi}
	\end{equation}
	i.e. Bob can apply $U_{12}$ to $|\tilde{\psi}\rangle$ then read the first qutrit to receive the original information $\ket{\psi}$! The process is what we mean by error correction: we encode our information into a large system $(\ket{\psi}\mapsto\ket{\tilde{\psi}})$ then use $U_{12}$ to recover the original information against the erasure of the third qutrit. Note that there exists permutation symmetry between qutrits in $\H_{code}$, which ensures that $U_{23},U_{31}$ also exist, i.e. Bob can recover the original information against the erasure of the first qutrit or the second qutrit as well.

	Formally, we consider a \emph{logical space} or \emph{code (sub)space} $\H_{code}$ which contains the \emph{true} degrees of freedom of the system, and a \emph{physical space} $\H_{phys}$ describing the system \emph{redundantly} such that $\H_{code}\subset \H_{phys}$, which we call the \emph{subspace description}. Equivalently, if we regard $\H_{code}$ and $ \H_{phys}$ as two \emph{independent} spaces, i.e. $\H_{code}$ is not regarded as the subspace of $ \H_{phys}$, then the encoding process is manifested by an isometric coding:
	\begin{equation}
		V:\H_{code}\rightarrow \H_{phys}
	\end{equation}
	which can induce a quantum channel from the code space to the physical space. We call it the \emph{encoding description}. In the case of the three-qutrit code, the isometric coding is given by
  \begin{equation}\label{threequtrits2}
  	\begin{aligned}
  		& |\widetilde{0}\rangle\mapsto\frac{1}{\sqrt{3}}(|000\rangle+|111\rangle+|222\rangle) \\
  		& |\widetilde{1}\rangle\mapsto\frac{1}{\sqrt{3}}(|012\rangle+|120\rangle+|201\rangle) \\
  		& |\widetilde{2}\rangle\mapsto\frac{1}{\sqrt{3}}(|021\rangle+|102\rangle+|210\rangle) .
  	\end{aligned}
  \end{equation}
  which one can compare with \eqref{threequtrits} described by the subspace description.

\begin{figure}[ht]
		\centering
		\includegraphics[width=\textwidth]{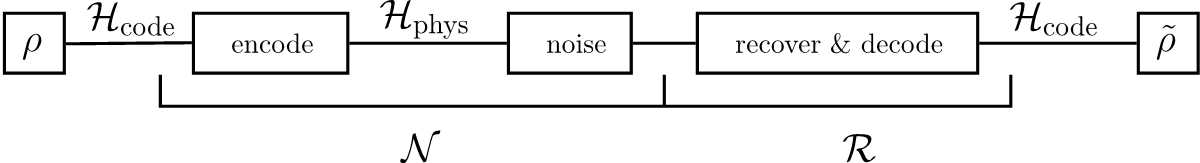}
		\caption{Extracted from \cite{Chatwin-Davies:2021nhs}. Quantum error correction can be realized by a series of quantum channels, where $\N$ represents encoding and noise, and $\R$ represents recovery and decoding.}
		\label{fig:QECasChannel}
	\end{figure}

	After encoding, the protected information will go through some latent errors or noise, and we still need to \emph{recover} and \emph{decode} the outcome to receive the information. The whole process is illustrated in Fig \ref{fig:QECasChannel} where we use a quantum channel $\N$ to represent encoding and noise, and another quantum channel $\R$, which is called the \emph{recovery channel}, to represent recovery and decoding. If the whole process is qualified to be a good QEC code, we expect that
	\begin{equation}\label{QECcriterion}
		\tilde{\rho}\equiv(\mathcal{\R} \circ \mathcal{N})(\rho) \approx \rho,\quad \forall\rho\in s(\H_{code})
	\end{equation}

 We say $\N$ is \emph{reversible} if $\R$ exists, and is \emph{exactly reversible} if the equality holds in \eqref{QECcriterion}. How do we know if a channel is exactly reversible or not? There is a theorem due to Petz \& Ohya \cite{ohya2004quantum} which can be demonstrated as follows,
 \begin{thm}\label{PetzOhya}
 	A quantum channel $\N:s(\H_{A})\rightarrow s(\H_{B})$ is exactly reversible if and only if 
 \begin{equation}
 	S(\rho|\sigma)=S(\N(\rho)|\N(\sigma)),\quad \forall \rho,\sigma\in s(\H_{A})
 \end{equation}
 Furthermore, the \emph{Petz map}:
 \begin{equation}\label{PetzMap}
 	\mathcal{P}_{\mathcal{\sigma},\mathcal{N}} ~ : ~ \gamma ~ \mapsto ~ \sigma^{1/2} \mathcal{N}^\dagger \left[ \mathcal{N}(\sigma)^{-1/2} \, \gamma \, \mathcal{N}(\sigma)^{-1/2}  \right] \sigma^{1/2},\quad \forall \gamma \in s(\H_{B})
 \end{equation} 
 is the recovery channel:
 \begin{equation}
 	(\mathcal{P}_{\mathcal{\sigma},\mathcal{N}} \circ \mathcal{N})(\rho) = \rho,\quad \forall \rho\in s(\H_{A})
 \end{equation}
 \end{thm}

Practically, it is very hard for a quantum channel to be reversible exactly, so we need a criterion for a channel to be reversible in an approximate sense. Due to theorem \ref{relentropymono} and theorem \ref{PetzOhya}, we say that $\N$ is only \emph{approximately} reversible if and only if
\begin{equation}\label{AQECcriterion}
	S(\rho|\sigma)-S(\N(\rho)|\N(\sigma))<\delta
\end{equation}
for some infinitesimal positive $\delta$. The procedure described in Fig \ref{fig:QECasChannel} together with \eqref{AQECcriterion} constitute the notion of \emph{approximate quantum error correction} (AQEC), see \cite{Junge_2018} and also earlier works \cite{barnum2000reversing,HIAI_2011,Mandayam_2012}.

\subsection{The reconstruction theorem}\label{SecRecThm}

The theorem in this section collects results from a series of work \cite{Almheiri:2014lwa,Dong:2016eik,Harlow:2016vwg,Kamal:2019skn}, and our setup is essentially the same as the theorem 3.1 in \cite{Harlow:2016vwg}. Additional to notations in section \ref{SecNotation}, we use $\mathcal{H}_{code}$ to denote the code space of the system and the ``tilde'' symbol is used for states in the code space, e.g. $\ket{\tilde{\psi}}\in \H_{code}$ and for operators acting on $\H_{code}$, e.g. $\tilde{O}\in\L(\H_{code})$.


\begin{thm}\label{RecThm}
	Given a finite-dimensional system $\H$ which is decomposed into two parts $\H=\H_{A}\otimes\H_{\bar{A}}$ and a code space $\H_{code}$ with condition $dim\H_{code}\leq\abs{A}$, the isometric encoding $V: \H_{code}\rightarrow \H$ induces a quantum channel:
	\begin{align}
		\N: ~&s(\H_{code})\rightarrow s(\H_A),\quad\tilde{\rho}\mapsto \rho_A\equiv\Tr_{\bar{A}}\left(V\tilde{\rho}V^\dagger\right)
	\end{align}
then the following statements are equivalent:
	\begin{enumerate}

		\item Given $\tilde{\rho}\in s(\H_{code})$, we have\footnote{Here we drop the familiar coefficient $\frac{1}{4G_N}$ because it can be regarded absorbed by the area term.}
		\begin{align}
			&S(\rho_A)=\mathcal{L}_A+S(\tilde{\rho})\label{rt1}
		\end{align}
		where $\mathcal{L}_A$ is some constant.
		
		\item
		
		Given two density matrices $\tilde{\rho},\tilde{\sigma}\in s(\H_{code})$, we have
		\begin{align}
			S(\rho_A|\sigma_A)&=S(\tilde{\rho}|\tilde{\sigma})
		\end{align}

		\item For any operator $\tilde{O}\in\mathcal{L}(\H_{code})$, there exists an operator $O_A\in\mathcal{L}(\H_A)$ such that
		\begin{equation}
			O_AV\ket{\tilde{\psi}}=V\tilde{O}\ket{\tilde{\psi}},\quad \forall \ket{\tilde{\psi}}\in \H_{code}
		\end{equation}

	\end{enumerate}

\end{thm}

The theorem is originally proved for reinterpreting finite-dimensional holographic models into a set of equivalent statements. In holographic theories \cite{Maldacena:1997re}, statement 1 and statement 2 correspond to the RT formula or the QES formula \cite{Ryu:2006bv,Hubeny:2007xt,Faulkner:2013ana,Engelhardt:2014gca}, and the JLMS formula \cite{Jafferis:2015del} respectively. The condition $dim\H_{code}\leq\abs{A}$ of the theorem says that if we expect the code space is \emph{reconstructable} from $A$, then $A$ must contain at least the same information of the code space such that the dimension of $A$ is at least equal to the dimension of the code space. By ``the code space is reconstructable from $A$'' we mean that if one manipulates some degrees of freedom on a \emph{known} region $A$, we are able to know how an \emph{unknown} region, which is the code space in this case, reacts. This statement indicates that there exists a \emph{dictionary} between operators on the two relevant regions, which is exactly the physical interpretation of statement 3 in the theorem. In the context of holography, this interprets the so-called GKP-W dictionary \cite{Witten:1998qj,Gubser:1998bc} which relates bulk operators and boundary operators.

In the paper, we do not need the holographic interpretation of the theorem because we expect that the island formula \eqref{islandFormula} describing the black hole evaporation holds generally, and we will use the theorem to model the black hole evaporation without holographic description in the next subsection. Essentially, the theorem itself is proved purely from the quantum information perspective and is not necessarily related to holography. Nevertheless to say, readers are encouraged to gain more insights from discussions involving holography. To this purpose, we give a proof of a slightly different version of the reconstruction theorem in appendix \ref{secProof}, which is more closely related to holographic models like AdS$_3$/CFT$_2$ correspondence. For a more accessible introduction, one can consult \cite{Harlow:2018fse,Pollack:2021yij}.

\section{Modeling black hole evaporation}\label{SecMain}

In this section we argue that the reconstruction theorem is applicable to the black hole evaporation. We first compare the theorem \ref{RecThm} with the island formula:
\begin{equation}\label{islandFormula3}
	S(\rho_{ R})=\text{min}\left\{\text{ext}_{X}\left[\frac{Area(X)}{4G}+S_{bulk}(\tilde{\rho}_{I\cup  R})\right]\right\}
\end{equation}  
then we find that when a non-vanishing $X$ is already specified, the island formula in the island phase, i.e. $S(\rho_{ R})=S^{island}_R= \frac{Area(X)}{4G}+S(\tilde{\rho}_{I\cup R})$ with a non-vanishing island $I$ after the Page time, has the same form as \eqref{rt1} in statement 1. Motivated by this observation, we can identify the region $R$ in the formula as the region $A$ in the theorem \ref{RecThm}. What follows is that, we can further regard the ``semi-classical'' space $\H^{sc}$ where $\tilde{\rho}_{I\cup  R}$ operates on as the code space $\H_{code}$, and the ``complete quantum gravity'' space $\H^{qg}$, which includes quantum gravitational effects, as the physical space $\H$ in the theorem \ref{RecThm}. 

To verify if the above identification works, we notice that the only condition for theorem \ref{RecThm} is $\dim\H_{code}\leq\abs{A}$, so we only need to verify the corresponding spaces satisfy the relation as well. We first notice that $\H^{qg}$ contains more information than $\H^{sc}$ because the latter is a semi-classical approximation of the former, i.e. $\H^{sc}\hookrightarrow \H^{qg}$. When referring to setup in Fig \ref{fig:island}, we notice that the spaces satisfy
\begin{equation}
	\H^{qg}_{R}\hookrightarrow\H^{qg},\quad \H^{sc}_{R}\hookrightarrow\H^{sc}=\H^{sc}_B\otimes\H^{sc}_{I\cup R}
\end{equation}
where $\H^{qg}_{R}$ denotes partial contribution from degrees of freedom in $R$ to the total space $\H^{qg}$ (likewise for $\H^{sc}_{R}$), and we need to clarify our notation for ``$I\cup R$'' which we use to denote that degrees of freedom in $I$ and degrees of freedom in $R$ are \emph{not independent}. The physical interpretation of this constraint is that the radiation of black hole is ``entangled'' with the island. An intuitive explanation is that when we discuss quantum theory in gravitational background, vacuum fluctuates even in the region around the black hole horizon, which looks ``normal'' as the usual spacetime when considering small enough region. If a pair of entangled particles emerges, with one of which appearing inside the horizon and the other one appearing outside, the inside particle will remain inside forever, which is believed to be the source of the island $I$, while the outside particle will radiate outwards and then finally be observed by observers in $R$. This correlation necessarily sets the number of degrees of freedom describing $I$ combined with $R$ to be exactly equal to that of $R$\footnote{We can regard such correlation between $I$ and $R$ as a \emph{coding relation} \cite{Basu:2022crn}.}:
\begin{equation}\label{eq1}
	\dim \H^{sc}_{I\cup R}=\dim \H^{sc}_{R}
\end{equation}

Besides, we only consider the \emph{non-gravitational} degrees of freedom within the combined space, which is due to the fact that the second term in the island formula is calculated in the semi-classical regime. i.e. it is computed in the curved but non-gravitational background. Therefore, the number of degrees of freedom describing $R$ in the semi-classical description is less than that in the complete quantum gravity description:
\begin{equation}\label{eq2}
	\dim \H^{sc}_{R}<\dim \H^{qg}_{R}
\end{equation}
then combining \eqref{eq1} and \eqref{eq2} implies
\begin{equation}
	\dim \H^{sc}_{I\cup R}<\dim \H^{qg}_{R}
\end{equation}
as required by the theorem \ref{RecThm} after making the following identifications:
\begin{equation}\label{ID}
	\begin{aligned}
		&\H\sim \H^{qg}, \quad \H_{code}\sim  \H_{I\cup R}^{sc},\quad  \H_A\sim \H^{qg}_R\\
		\Rightarrow &\dim\H_{code}\leq\abs{A}~~\sim~~ 	\dim \H^{sc}_{I\cup R}<\dim \H^{qg}_{R}
	\end{aligned}
\end{equation}

After verifying that the theorem \ref{RecThm} can be applied to the island formula, we now have statement 1 in the theorem after \eqref{ID}, such that the following statements hold simultaneously according to the theorem:
	\begin{itemize}
		\item (Statement A) Given $\tilde{\rho},\tilde{\sigma}$ on $\H^{sc}$: $S(\rho_{R}|\sigma_{R})=S(\tilde{\rho}_{I\cup R}|\tilde{\sigma}_{I\cup R})$
		
		\item (Statement B) Given $\tilde{O}_{I\cup R}\in\L(\H^{sc}_{I\cup R})$: $\exists O_R\Rightarrow O_R V|\tilde{\phi}\rangle=V\tilde{O}_{I\cup R}|\tilde{\phi}\rangle$
		
	\end{itemize}
	
To convince the readers and also as a calculation warm-up, we now show how to derive the statement A from the island formula \eqref{islandFormula3} in the island phase, then the statement B is straightforward from applying the theorem \ref{RecThm}. The derivation basically follows \cite{Dong:2016eik,Jafferis:2015del}. We start with
\begin{equation}\label{eqS1}
	S(\sigma_R)=S(\tilde{\sigma}_{I\cup R})+\L_X=S(\tilde{\sigma}_{I\cup R})+(\Tr\tilde{\sigma})\L_X
\end{equation}
where the constant $\L_X$ denotes the area term and we use the fact that a density matrix has unital trace, i.e. $\Tr\tilde{\sigma}=1$. We arbitrarily perturbate $\tilde{\sigma}$ to first order:
\begin{equation}\label{eqS2}
	\Tr_{R}\left(\delta \sigma_{R} K^{\sigma}_{R}\right)=\Tr_{I\cup R}\left(\delta \tilde{\sigma}_{I\cup R} \tilde{K}^{\sigma}_{I\cup R}\right)+(\Tr\delta\tilde{\sigma})\L_X
\end{equation}
where $K^{\sigma}_{R}$ is the modular Hamiltonian for the density matrix $\sigma_R$ defined by $K^{\sigma}_{R}\equiv -\log \sigma_R$ and likewise for $\tilde{K}^{\sigma}_{I\cup R}$, and we have used the fact that perturbation of the von Neumann entropy $S(\rho)$ with respect to the density matrix is given by
\begin{equation}\label{varS}
	\delta S(\rho)\equiv S(\rho+\delta \rho)-S(\rho)\approx-\Tr\left(\delta\rho\log\rho\right)\equiv \Tr\left(\delta\rho K^\rho\right)
\end{equation} 
such that the LHS of \eqref{eqS1} becomes the LHS of \eqref{eqS2} and likewise for the first terms of the RHS of \eqref{eqS1} and \eqref{eqS2}. 

Both sides of \eqref{eqS2} are now linear in the infinitesimal entries of $\delta\tilde{\sigma}$, which can be integrated from zero to $\tilde{\rho}$:
\begin{equation}\label{eqK1}
	\Tr_{R}\left(\rho_{R} K^{\sigma}_{R}\right)=\Tr_{I\cup R}\left( \tilde{\rho}_{I\cup R} \tilde{K}^{\sigma}_{I\cup R}\right)+\L_X
\end{equation}
where the term ``integration from zero to $\tilde{\rho}$'' is to denote integrating the infinitesimal entries of $\delta\tilde{\sigma}$ to be the entries of $\tilde{\rho}$. Now \eqref{eqK1} implies
\begin{equation}
	\begin{aligned}
		S(\tilde{\rho}_{I\cup R}|\tilde{\sigma}_{I\cup R})-S(\rho_{R}|\sigma_{R})&=-\Tr_{R}\left(\rho_R K^\sigma_R\right)+\Tr_{I\cup R}\left(\tilde{\rho}_{I\cup R} \tilde{K}^\sigma_{I\cup R}\right)+S(\rho_R)-S(\tilde{\rho}_{I\cup R})\\
		&=0
	\end{aligned}
\end{equation}
where the first equality makes use of the definition of the relative entropy and the second one uses \eqref{eqS1} and \eqref{eqK1}, completing our derivation to the first statement.

The statement A is exactly the criterion for the quantum channel $\N:s(\H^{sc}_{I\cup R})\rightarrow s(\H^{qg}_{R})$ to be exactly reversible in the language of AQEC. The statement B says that the island $I$ is \emph{reconstructable} from $R$. Combining the two we find that observers in $R$ do ``observe'' the island in this exactly-reversible case. But what if $\N$ is \emph{not} exactly reversible? It is indeed the case for the island formula before the Page time. As we have discussed, the following two cases are possible when $\N$ is not exactly reversible:
	\begin{itemize}
		 
		\item If $S(\tilde{\rho}_{I\cup R}|\tilde{\sigma}_{I\cup R})-S(\rho_{R}|\sigma_{R})< \delta$, then there exist an operator $O_R$ such that $O_R V|\tilde{\phi}\rangle\approx V\tilde{O}_{I\cup R}|\tilde{\phi}\rangle$, i.e. observers in $R$ \emph{approximately} observe the island.

		\item If $S(\tilde{\rho}_{I\cup R}|\tilde{\sigma}_{I\cup R})-S(\rho_{R}|\sigma_{R})<\!\!\!\!\!/ \delta$, then we have $O_R V|\tilde{\phi}\rangle\neq V\tilde{O}_{I\cup R}|\tilde{\phi}\rangle$, i.e. observers in $R$ do \emph{not} observe the island.

	\end{itemize}
We can see that the difference of relative entropies $S(\tilde{\rho}_{I\cup R}|\tilde{\sigma}_{I\cup R})-S(\rho_{R}|\sigma_{R})$ can be used to characterize the appearance of $O_R$, i.e. a phase transition from ``phase without the existence of $O_R$'' to ``phase with the existence of $O_R$'', at the critical point $S(\tilde{\rho}_{I\cup R}|\tilde{\sigma}_{I\cup R})-S(\rho_{R}|\sigma_{R})= \delta$. Could it describe the Page transition? If true, we expect that the black hole evaporation satisfies
	\begin{equation}\label{Page Transition}
		S(\tilde{\rho}_{I\cup R}(t)|\tilde{\sigma}_{I\cup R}(t))-S(\rho_{R}(t)|\sigma_{R}(t))\begin{cases}
			>\delta,  & \text{if $t<t_{Page}$} \\
			=\delta, & \text{if $t=t_{Page}$}\\
			\leq\delta, & \text{if $t>t_{Page}$}
		\end{cases}
	\end{equation}
	for some threshold value $\delta$.

\subsection{AQEC as a probe for the Page transition}

\begin{figure}[ht]
	\centering
	\includegraphics[scale=0.25]{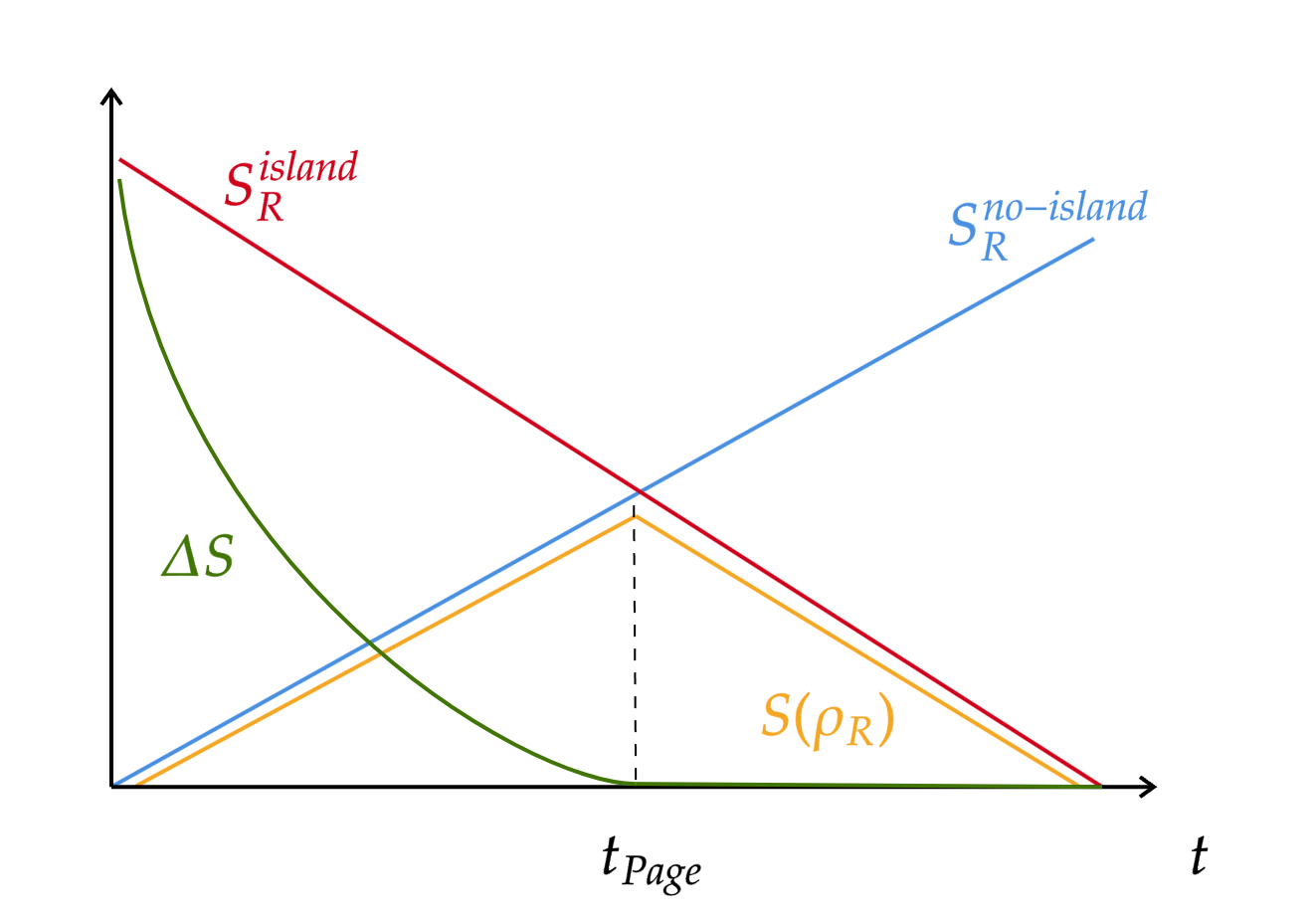}
	\caption{The vanishing island solution $S^{no-island}_R$ and the non-vanishing island solution $S^{island}_R$ are presented by the blue line and the red line respectively. The entropy of radiation $S(\rho_{ R})$ follows the Page curve presented by the orange line. $\Delta S$, defined by the difference between $S^{island}_R$ and $S(\rho_{ R})$, is non-vanishing and monotonically decreasing before $t_{Page}$, and then vanishes after $t_{Page}$.}
	\label{PageCurve}
\end{figure}

Now we are ready to model black hole evaporation by proving that \eqref{Page Transition} holds. We rewrite the island formula as
\begin{equation}
	S(\rho_R)=\min\{ S^{island}_R,S^{no-island}_R\}
\end{equation}
where $ S^{island}_R= S(\tilde{\rho}_{I\cup R})+\L_X$ and here $I$ denotes the non-vanishing island solution. We further define 
\begin{equation}\label{DelS}
	\Delta S\equiv S^{island}_R-S(\rho_R)
\end{equation}
Note that \eqref{DelS} is a function of time $t$ via the state $\rho_{ }(t)$ describing the black hole during the evaporation process, and \eqref{DelS} is non-vanishing and monotonically decreasing before $t_{Page}$, then vanishes after $t_{Page}$: 
\begin{equation}\label{DelSprop}
	\Delta S(t)\begin{cases}
		>0,  & \text{if $t<t_{Page}$} \\
		=0, & \text{if $t\geq t_{Page}$}
	\end{cases},\quad \frac{dS(t)}{dt}\begin{cases}
	<0,  & \text{if $t<t_{Page}$} \\
	=0, & \text{if $t\geq t_{Page}$}
	\end{cases}
\end{equation}
as illustrated in Fig \ref{PageCurve}. Now our strategy is to rewrite
	\begin{equation}
		\Delta S_{rel}(\tilde{\rho}|\tilde{\sigma})\equiv S(\tilde{\rho}_{I\cup R}|\tilde{\sigma}_{I\cup R})-S(\rho_{R}|\sigma_{R})
	\end{equation}
in terms of $\Delta S$ such that the property of $\Delta S$ can be adopted. According to the definition \eqref{DelS} of $\Delta S$:
\begin{equation}\label{eqS}
	S(\rho_R)=S(\tilde{\rho}_{I\cup R})+\L_X-\Delta S(\tilde{\rho})
\end{equation}
from which we regard $\Delta S$ as the perturbation of the entropy of radiation around the non-vanishing island solution. Since $\Delta S$ vanishes after $t_{Page}$, $\Delta S_{rel}$ also vanishes according to our previous island phase calculation, then we only need to examine if 
\begin{equation}\label{DelSrel0}
	\lim_{t\rightarrow t_{Page}^-}\Delta S_{rel}\rightarrow 0
\end{equation}
before the Page time. To verify, we start with
\begin{equation}
	S(\sigma_R)=S(\tilde{\sigma}_{I\cup R})+\L_X-\Delta S(\tilde{\sigma})
\end{equation}
and we perturbate $\tilde{\sigma}$ to first order:
\begin{equation}
	\Tr_{R}\left(\delta \sigma_{R} K^{\sigma}_{R}\right)=\Tr_{I\cup R}\left(\delta \tilde{\sigma}_{I\cup R} \tilde{K}^{\sigma}_{I\cup R}\right)+\Tr\delta\tilde{\sigma}\L_X-\delta\Delta S(\tilde{\sigma})
\end{equation}
where $\delta\Delta S(\tilde{\sigma})\equiv \Delta S(\tilde{\sigma}+\delta\tilde{\sigma})-\Delta S(\tilde{\sigma})$. Integrating $\delta\tilde{\sigma}$ from zero to $\tilde{\rho}$:
	\begin{equation}\label{eqK}
		\Tr_{R}\left(\rho_{R} K^{\sigma}_{R}\right)=\Tr_{I\cup R}\left( \tilde{\rho}_{I\cup R} \tilde{K}^{\sigma}_{I\cup R}\right)+\L_X-\int_{0}^{\tilde{\rho}}\delta\Delta S(\tilde{\sigma})
	\end{equation}
and then expanding $\Delta S_{rel}(\tilde{\rho}|\tilde{\sigma})$:
	\begin{equation}
		\begin{aligned}
			\Delta S_{rel}(\tilde{\rho}|\tilde{\sigma})&=-\Tr_{R}\left(\rho_R K^\sigma_R\right)+\Tr_{I\cup R}\left(\tilde{\rho}_{I\cup R} \tilde{K}^\sigma_{I\cup R}\right)+S(\rho_R)-S(\tilde{\rho}_{I\cup R})\\
			&=\int_{0}^{\tilde{\rho}}\delta\Delta S(\tilde{\sigma})-\Delta S(\tilde{\rho})
		\end{aligned}
	\end{equation}
	where the second equality uses \eqref{eqS} and \eqref{eqK}. The final expression is positive which is guaranteed by the theorem \ref{relentropymono}, and can be regarded as a sum of $\Delta S$'s with different variables because basically integration is an infinite sum of the integrand, with each terms vanishing at $t_{Page}$ according to \eqref{DelSprop}.	We therefore conclude that the criterion \eqref{DelSrel0} holds indeed.

 \subsection{Threshold of the Page transition}

	In fact, different states result in different geometries in gravitational theory, hence we have different geometric configurations for $\tilde{\rho}$ and $\tilde{\sigma}$ (i.e. $X^{\{\rho\}}\neq X^{\{\sigma\}}$ in Fig \ref{fig:island} where the superscript denotes which density matrix is associated) which may result in the non-vanishing threshold value. During the verification in the last section we implicitly assume that the Hilbert space factorization $\H^{sc}=\H^{sc}_B\otimes\H^{sc}_{I\cup R}$ is the same for different density matrices, which is not necessarily true. Meanwhile, if two density matrices give two hugely different Hilbert space factorizations will make our derivation in the last section no longer valid.
	
	Therefore, one additional constraint should be manually added: 
	\begin{itemize}
		\item \emph{Two different density matrices which we choose to calculate the relative entropy should correspond to black holes with the same macroscopic quantities}. 
	\end{itemize}
	
	To be specific, when we calculate $\Delta_{rel}S$, we can only choose density matrices from $s(\H_{macro,\{\ell\}})\subset s(\H^{sc})$ where the subscript ``macro'' denotes that the states within give the same macroscopic quantities and ``$\{\ell\}$'' collectively specifies all possible macroscopic quantities. The constraint is natural because for observers localized in $R$, they should only observe the macroscopic quantities while the microscopic details are still hidden inside the horizon. The constraint results in the fact that the two density matrices we choose should approximately give the same radius of black hole horizon, which further implies that $X^{\{\rho\}}\approx X^{\{\sigma\}}$ as the QES is slightly deviated from the horizon. The other consequence is that the black holes for two density matrices should be formed in the same time. As a counter example, two same-size black holes with one of which being in the early time of evaporation while the other being in the later time of evaporation have vastly different entropies of radiation, because the former do not have island contribution while the latter does.

	After carefully specifying which configuration we are considering in the above calculation, one can perform a similar procedure in the last subsection to show that after $t_{Page}$ we have
	\begin{equation}
		\Delta S_{rel}(\tilde{\rho}|\tilde{\sigma})\equiv S(\tilde{\rho}^{\{\rho\}}_{I\cup R}|\tilde{\sigma}^{\{\sigma\}}_{I\cup R})-S(\rho_{R}^{\{\rho\}}|\sigma_{R}^{\{\sigma\}})=\delta_1(\tilde{\rho},\tilde{\sigma})+\delta_2(\tilde{\rho},\tilde{\sigma}),\quad \tilde{\rho},\tilde{\sigma}\in s(\H_{macro,\{\ell\}})
	\end{equation}
	where 
	\begin{align}
		&\delta_1(\tilde{\rho},\tilde{\sigma})\equiv S(\tilde{\rho}_{I\cup  R}^{\{\rho\}}|\tilde{\sigma}_{I\cup  R}^{\{\sigma\}})-S(\tilde{\rho}_{I\cup  R}^{\{\sigma\}}|\tilde{\sigma}_{I\cup  R}^{\{\sigma\}}),\\
		&\delta_2(\tilde{\rho},\tilde{\sigma})\equiv \left(S(\tilde{\rho}_{I\cup  R}^{\{\rho\}})+\L_X^{\{\rho\}}\right)-\left(S(\tilde{\rho}_{I\cup  R}^{\{\sigma\}})+\L_X^{\{\sigma\}}\right)
	\end{align}
According to our constraint that $X^{\{\rho\}}\approx X^{\{\sigma\}}$, $\delta_1(\tilde{\rho},\tilde{\sigma})$ is an infinitesimal value. As for $\delta_2(\tilde{\rho},\tilde{\sigma})$, we recognize that the term inside round bracket is exactly what is extremized by QES in \eqref{islandFormula}. Therefore, $\delta_2(\tilde{\rho},\tilde{\sigma})$ is only non-vanishing in second order of configuration perturbation. The threshold value can now be chosen by maximizing $\delta_1(\tilde{\rho},\tilde{\sigma})+\delta_2(\tilde{\rho},\tilde{\sigma})$ among all $\tilde{\rho},\tilde{\sigma}\in s(\H_{macro,\{\ell\}}) $ and among all Hilbert spaces with different macroscopic quantities:
\begin{equation}
	\delta\equiv \text{max}_{\tilde{\rho},\tilde{\sigma},\{\ell\}}\left(\delta_1(\tilde{\rho},\tilde{\sigma})+\delta_2(\tilde{\rho},\tilde{\sigma})\right),\quad \tilde{\rho},\tilde{\sigma}\in s(\H_{macro,\{\ell\}})
\end{equation}
which concludes our proof for \eqref{Page Transition} to be held.

\section{Discussions}\label{SecDiscussions}

In the paper, we have verified that the black hole evaporation does satisfy the criterion \eqref{AQECcriterion} of AQEC after the Page time hence the island emerges from the viewpoint of outside observers, and we have also calculated the threshold value of AQEC criterion for black hole evaporation. There are two aspects we would like to emphasize:
\begin{itemize}
	\item Gravity plays ``almost'' no role.

	\item Our definition of the Page transition requires the notion of \emph{local} observers.
\end{itemize}

The first statement may be a bit misleading. By it we mean that gravity only manifests itself in two ways. The first one gives the area term in \eqref{islandFormula} while the statement 1 in reconstruction theorem \ref{RecThm} allows arbitrary value of $\L_A$\footnote{It is pointed out in \cite{Akers:2018fow} that $\L_A$ corresponds to the entropy of a maximally mixed state in gauge theories, which applies to gravity with diffeomorphism being the corresponding gauge symmetry.}, i.e. it can vanish, which corresponds to the non-gravitational cases. The other one is that the Hilbert space of $I$ combined with $R$ has both gravitational and non-gravitational contributions while the second term in the island formula \eqref{islandFormula} only consider the non-gravitational part. Recall that in using the reconstruction theorem \ref{RecThm} we need to ensure that the dimension of the code space should be less than or \emph{equal} to $\abs{A}$, and the latter constraint that only the non-gravitational part is considered is \emph{not} necessary for the theorem to be applicable to the black hole evaporation, because the constraint \eqref{eq1} between $I$ and $R$ already ensures that the theorem \ref{RecThm} works.

The second statement requires the locality of observers. By ``locality'' we mean that observers are only able to access degrees of freedom of some certain subregions, while any other subregion which is space-like separated to the accessible subregion is blind to those observers. For example, provided a density matrix $\rho$ describing the whole system in the setup of Fig \ref{fig:island}, observers located in $R$ can only read or manipulate degrees of freedom in $R$. Therefore, when observers located in $R$ measure the state, they only observe the reduce density matrix $\rho_R\equiv \Tr_{\bar{R}}\rho$ rather than $\rho$. A consequence is that when observers in $R$ talk about von Neuman entropy, they can only refer to $S(\rho_{ R})$ rather than $S(\rho)$ of the total state, and then according to the reconstruction theorem \ref{RecThm}, observers located in $R$ only observe the existence of island when an island is needed to correctly calculate the entropy $S(\rho_{ R})$. In the case of black hole evaporation, what we focus on is the entropy of radiation, which is only accessible to observers in $R$. Besides, the requirement for locality is manifest when we try to use the statement 3 in the reconstruction theorem \ref{RecThm}, where in the case of black hole evaporation we have $\exists O_R\Rightarrow O_R V|\tilde{\phi}\rangle=V\tilde{O}_{I\cup R}|\tilde{\phi}\rangle$ for any given $\tilde{O}_{I\cup R}\in\L(\H^{sc}_{I\cup R})$. The reconstruction from operators on $R$ requires manipulation of degrees of freedom on $R$ hence the locality condition is necessary.

Motivated by these arguments, we make the following generalization,

\begin{Proposal}\label{Proposal}
	For a \emph{general} system $\H_I\otimes \H_{B}\otimes \H_{R}$ as in Fig \ref{fig:island}, if applying some parameter-dependent constraints on $I$ from $R$ results in 
	\begin{equation}\label{eqproposal}
		S(\tilde{\rho}_{I\cup R}(s)|\tilde{\sigma}_{I\cup R}(s))-S(\rho_{R}(s)|\sigma_{R}(s))\begin{cases}
			>\delta,  & \text{if $s<s_{p}$} \\
			=\delta, & \text{if $s=s_{p}$}\\
			\leq\delta, & \text{if $s>s_{p}$}
		\end{cases}
	\end{equation}
	for some threshold value $\delta$, then the system possesses a transition which accounts for the emergence of the island at the critical point $s=s_p$, from the viewpoint of observers in $R$.
\end{Proposal}

The parameter $s$ characterizes the evolution of the system, and in the case of black hole evaporation it corresponds to time $t$. The parameter-dependent constraints on $I$ from $R$ are required to set $\abs{I\cup R}\leq\abs{R}$ as $s$ evolves, with the condition that the less-than sign only holds when other constraints are manually added, e.g. issues involving gravity. When the condition \eqref{eqproposal} is satisfied due to such constraints, then according to the reconstruction theorem \ref{RecThm} we have $\exists O_R\Rightarrow O_R V|\tilde{\phi}\rangle\approx V\tilde{O}_{I\cup R}|\tilde{\phi}\rangle$ for any given $\tilde{O}_{I\cup R}\in\L(\H_{I\cup R})$ only after the critical point $s=s_p$. Therefore, the system in this case, with or without gravity, exhibits a transition from a vanishing island state to a non-vanishing island state, which can be measured by calculating the von Neumann entropy over $R$.

It is explained that the constraint on $I$ from $R$ is the origin for the island in \cite{Basu:2022crn,Basu:2023wmv}, where authors consider a \emph{coding relation} $\ket{i}_R\Rightarrow \ket{f(i)}_I$ which projects out certain states in the original Hilbert space, resulting in \emph{Hilbert space reduction}\footnote{In the language of QEC, such Hilbert space reduction is exactly the encoding $V: \H_{code}\rightarrow\H$ with $\H_{code}$ being the Hilbert space after reduction and $V$ being the embedding mapping.} which claims to give a general version of the island formula. To be specific, when one actually computes the gravitational path integral by using the Replica trick \cite{Lewkowycz:2013nqa} to calculate the black hole entropy of radiation, the constraint on $I$ from $R$ sets the boundary conditions for $I$ when the boundary conditions for $R$ are set. Therefore, additional gluing of $I$ over replicas gives a twist operator asserted at the location of $X$, which contributes the area term to the island formula. However, the claim only discussed how the island modifies the entropy over a \emph{single} time-slice. The mechanism of a transition from non-island phase to island phase during the evolution of the system was left as an unsolved problem, and authors in \cite{Basu:2022crn,Basu:2023wmv} suspected that there should exist more yet-to-known constraints to make the mechanism manifest. Our proposal \ref{Proposal} solves this problem and the condition \eqref{eqproposal} is exactly the requirement for such transition to exist.

\acknowledgments

The author would like to thank Zhenbin Yang, Qiang Wen, Mingshuai Xu for helpful discussions. This work was supported by SEU Innovation Capability Enhancement Plan for Doctoral Students (Grant No. CXJH\_SEU 24137).

\appendix

\section{Proof of the reconstruction theorem}\label{secProof}

In this appendix we prove a slightly different version of the reconstruction theorem in section \ref{SecRecThm}. Besides that, instead of proving the three statements in theorem \ref{RecThm} to be equivalent, we also enlarge the theorem to include five statements. For simplicity in notations, we will use the subspace description that the code space is regarded as the subspace of the physical space, i.e. $\H_{code}\subset \H_{phys}$.

\begin{figure}[h]
	\centering
	\includegraphics[width=0.5\textwidth]{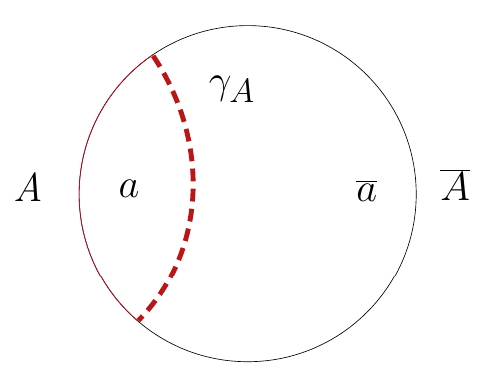}
	\caption{A time slice for AdS$_3$/CFT$_2$. The boundary region is divided in two parts: subregion $A$ represented by red solid line, and its complementary $\bar{A}$ represented by black solid line. RT curve $\gamma_A$ is represented by the red dashed line, and the entanglement wedges of $A$ and $\bar{A}$ are denoted by $a$ and $\bar{a}$ respectively.} 
	\label{SubDual}
\end{figure}

\begin{thm}
	Given a finite-dimensional system $\H$ which is decomposed into two parts $\H=\H_{A}\otimes\H_{\bar{A}}$, and a code subspace $\H_{code}\subseteq \H$ which is decomposed as $\H_{code}=\H_a\otimes\H_{\bar{a}}$ with conditions $\abs{a}\leq\abs{A}$ and $\abs{\bar{a}}\leq\abs{\bar{A}}$, then we can make the following decomposition $\mathcal{H}_A=\H_{A_1}\otimes\H_{A_2}\oplus \H_{A_3}$ and $\mathcal{H}_{\bar{A}}=\H_{\bar{A}_1}\otimes\H_{\bar{A}_2}\oplus\H_{\bar{A}_3}$ with $\abs{a}=\abs{A_1}$ and $\abs{\bar{a}}=\abs{\bar{A}_1}$. We write the orthonormal basis of $\H_a$ as $\ket{\tilde{i}}_a, i=1,\dots, |a|$ and $\ket{\tilde{j}}_{\bar{a}}, j=1,\dots, |\bar{a}|$ for $\H_{\bar{a}}$. Similarly, we use $\ket{i}_{A_1},\ket{j}_{\bar{A}_1}$ for orthonormal basis of $\H_{A_1},\H_{\bar{A}_1}$ respectively. The following statements are equivalent:
	\begin{enumerate}
		
		\item Define $\ket{\tilde{ij}}\equiv \ket{\tilde{i}}_a\ket{\tilde{j}}_{\bar{a}}$, there exists a unitary operator $U_A\in \mathcal{L}(\H_{A})$ and a set of orthonormal states $\ket{\chi_j}_{A_2\bar{A}}\in\H_{A_2\bar{A}}$ such that 
		\begin{equation}\label{ijtilde1}
			\ket{\tilde{ij}}=U_A\left(\ket{i}_{A_1}\ket{\chi_j}_{A_2\bar{A}}\right)
		\end{equation}
		
		Similarly, there exists a unitary operator $U_{\bar{A}}\in \mathcal{L}(\H_{\bar{A}})$ and a set of orthonormal states $\ket{\bar{\chi}_i}_{\bar{A}_2 A}\in\H_{\bar{A}_2 A}$ such that 
		\begin{equation}\label{ijtilde2}
			\ket{\tilde{ij}}=U_{\bar{A}}\left(\ket{j}_{\bar{A}_1}\ket{\bar{\chi}_i}_{\bar{A}_2 A}\right)
		\end{equation}

		\item Define $\tilde{\rho}\equiv\sum_{i,j}p_{ij}\ket{\tilde{ij}}\bra{\tilde{ij}}$ as the density matrix on $\H_{code}$ where $p_{ij}$ satisfies $\sum_{ij}p_{ij}=1$, then for $\tilde{\rho}_A\equiv\Tr_{\bar{A}}\tilde{\rho}$ and $\tilde{\rho}_a\equiv\Tr_{\bar{a}}\tilde{\rho}$, we have the following relation
		\begin{align}\label{RT1}
			S(\tilde{\rho}_A)=\mathcal{L}_A+S(\tilde{\rho}_a)
		\end{align}
		Likewise for $\tilde{\rho}_{\bar{A}}\equiv\Tr_{A}\tilde{\rho}$ and $\tilde{\rho}_{\bar{a}}\equiv\Tr_{a}\tilde{\rho}$, we have
		\begin{equation}\label{RT2}
			S(\tilde{\rho}_{\bar{A}})=\mathcal{L}_{\bar{A}}+S(\tilde{\rho}_{\bar{a}})
		\end{equation}
		and more importantly,
		\begin{equation}
			\mathcal{L}_A=\mathcal{L}_{\bar{A}}
		\end{equation}

		\item
		
		Given two density matrices $\tilde{\rho},\tilde{\sigma}$ on $\H_{code}$, their reduced density matrices $\tilde{\rho}_A\equiv\Tr_{\bar{A}}\tilde{\rho},~\tilde{\rho}_a\equiv\Tr_{\bar{a}}\tilde{\rho},~\tilde{\sigma}_A\equiv\Tr_{\bar{A}}\tilde{\sigma},~\tilde{\sigma}_a\equiv\Tr_{\bar{a}}\tilde{\sigma}$ satisfy
		\begin{align}\label{relSeq1}
			S(\tilde{\rho}_A|\tilde{\sigma}_A)&=S(\tilde{\rho}_a|\tilde{\sigma}_a)
		\end{align}
		and likewise for $\tilde{\rho}_{\bar{A}}\equiv\Tr_A\tilde{\rho},~\tilde{\rho}_{\bar{a}}\equiv\Tr_a\tilde{\rho},~\tilde{\sigma}_{\bar{A}}\equiv\Tr_A\tilde{\sigma},~\tilde{\sigma}_{\bar{a}}\equiv\Tr_a\tilde{\sigma}$,
		\begin{align}\label{relSeq2}
			S(\tilde{\rho}_{\bar{A}}|\tilde{\sigma}_{\bar{A}})&=S(\tilde{\rho}_{\bar{a}}|\tilde{\sigma}_{\bar{a}})
		\end{align}

		\item Given any state $\ket{\tilde{\phi}}\in \H_{code}$ and an operator $\tilde{O}_a\in\mathcal{L}(\H_a)$, we have
		\begin{equation}\label{ComO1}
			\bra{\tilde{\phi}}[\tilde{O}_a,X_{\bar{A}}]\ket{\tilde{\phi}}=0,\quad \forall X_{\bar{A}}\in\mathcal{L}(\H_{\bar{A}})
		\end{equation}
		Likewise for an operator $\tilde{O}_{\bar{a}}\in\mathcal{L}(\H_{\bar{a}})$,
		\begin{equation}\label{ComO2}
			\bra{\tilde{\phi}}[\tilde{O}_{\bar{a}},X_{A}]\ket{\tilde{\phi}}=0,\quad \forall X_{A}\in \mathcal{L}(\H_{A})
		\end{equation}
		
		\item For any operator $\tilde{O}_a\in\mathcal{L}(\H_a)$, there exists an operator $O_A\in\mathcal{L}(\H_A)$ such that
		\begin{equation}\label{correct1}
			O_A\ket{\tilde{\psi}}=\tilde{O}_a\ket{\tilde{\psi}},\quad \forall \ket{\tilde{\psi}}\in \H_{code}
		\end{equation}
		Likewise for any operator $\tilde{O}_{\bar{a}}\in\mathcal{L}(\H_{\bar{a}})$, there exists an operator $O_{\bar{A}}\in\mathcal{L}(\H_{\bar{A}})$ such that
		\begin{equation}\label{correct2}
			O_{\bar{A}}\ket{\tilde{\psi}}=\tilde{O}_{\bar{a}}\ket{\tilde{\psi}},\quad \forall \ket{\tilde{\psi}}\in \H_{code}
		\end{equation}

	\end{enumerate}

\end{thm}
Comparing to the theorem \ref{RecThm}, we can see that this theorem further divides the code subspace into two parts: $a,\bar{a}$, with $a$ being reconstructable from $A$ and $\bar{a}$ being reconstructable from $\bar{A}$. In describing holography as in Fig \ref{SubDual}, $\H_{code}$ is the Hilbert space of AdS$_3$ and $\H$ is the Hilbert space of CFT$_2$. If the CFT$_2$ is divided into two subregions: $A,\bar{A}$, the so-called \emph{subregion duality} tells us that there exists a RT curve $\gamma_A$ which divides the bulk region into two which correspond to the entanglement wedges of boundary subregions, denoted by $a,\bar{a}$. Furthermore, $a,\bar{a}$ are reconstructable from $A,\bar{A}$ respectively.

The additional statement 1 in the theorem is analogous to \eqref{trans3q} in the three-qutrit example, which reminds us that we are doing QEC. The additional statement 4  also have its holographic interpretation, which is related to a puzzle called radial commutativity. Radial commutativity states that all local boundary operators should commute with bulk operators, and it is a puzzle because the statement seems to imply that all bulk operators should be trivial according to the time slice axiom \cite{Harlow:2018fse} which is a variational version of Schur's lemma. However, this implication only holds when the local boundary operators and the bulk operators act on the \emph{same} Hilbert space. As implicitly indicated in the statement 4, the commutation relations for bulk operators hold for the subspace $\H_{code}$ while the boundary operators are defined on $\H$, hence solving the puzzle.

Our proof will be given as follows: $(1)\Rightarrow(2)\Rightarrow(3)\Rightarrow(4)\Rightarrow(1)$ which put the first four statements in an equal footing, then we prove $(1)\Rightarrow(5)\Rightarrow(4)$ to include statement $(5)$. The proof is based on \cite{Harlow:2016vwg,Almheiri_2015,Dong:2016eik}. One can analogously prove the theorem \ref{RecThm} by considering only the ``$a$'' part of the code space.

\begin{proof}
	\begin{itemize}
		\item $(1)\Rightarrow(2)$: 
		Given \eqref{ijtilde1} and \eqref{ijtilde2}, acting $U^\dagger_A, U^\dagger_{\bar{A}}$ on $\ket{\tilde{ij}}$ gives 
		\begin{equation}
			U^\dagger_AU^\dagger_{\bar{A}}\ket{\tilde{ij}}=\ket{i}_{A_1}\left(U^\dagger_{\bar{A}}\ket{\chi_j}_{A_2\bar{A}}\right)=\ket{j}_{\bar{A}_1}\left(U^\dagger_A\ket{\bar{\chi}_i}_{\bar{A}_2 A}\right)
		\end{equation}
		Compare the last two results we find that there must exists $\ket{\chi}_{A_2\bar{A}_2},\ket{\bar{\chi}}_{A_2\bar{A}_2}$ which satisfy
		\begin{equation}
			U^\dagger_{\bar{A}}\ket{\chi_j}_{A_2\bar{A}}=\ket{j}_{\bar{A}_1}\ket{\chi}_{A_2\bar{A}_2},\quad U^\dagger_A\ket{\bar{\chi}_i}_{\bar{A}_2 A}=\ket{i}_{A_1}\ket{\bar{\chi}}_{A_2\bar{A}_2}
		\end{equation}
		which further implies
		\begin{equation}
			U^\dagger_AU^\dagger_{\bar{A}}\ket{\tilde{ij}}=\ket{i}_{A_1}\ket{j}_{\bar{A}_1}\ket{\chi}_{A_2\bar{A}_2}=\ket{j}_{\bar{A}_1}\ket{i}_{A_1}\ket{\bar{\chi}}_{A_2\bar{A}_2}
		\end{equation}
		i.e. in fact we have $\ket{\chi}_{A_2\bar{A}_2}=\ket{\bar{\chi}}_{A_2\bar{A}_2}$ such that \eqref{ijtilde1} can be simplified as
		\begin{equation}
			\ket{\tilde{ij}}=U_AU_{\bar{A}}\left(\ket{ij}_{A_1\bar{A}_1}\ket{\chi}_{A_2\bar{A}_2}\right)
		\end{equation}
		where $\ket{ij}_{A_1\bar{A}_1}\equiv\ket{i}_{A_1}\ket{j}_{\bar{A}_1}$, then for $\tilde{\rho}\equiv\sum_{i,j}p_{ij}\ket{\tilde{ij}}\bra{\tilde{ij}}$ we have
		\begin{equation}
			\tilde{\rho}=U_AU_{\bar{A}}\left(\rho_{A_1\bar{A}_1}\otimes \chi_{A_2\bar{A}_2}\right)U^\dagger_AU^\dagger_{\bar{A}}
		\end{equation}
		where we define
		\begin{equation}\label{chi}
			\rho_{A_1\bar{A}_1}\equiv \sum_{i,j}p_{ij}\ket{ij}_{A_1\bar{A}_1}\bra{ij}_{A_1\bar{A}_1},\quad \chi_{A_2\bar{A}_2}\equiv\ket{\chi}_{A_2\bar{A}_2}\bra{\chi}_{A_2\bar{A}_2}
		\end{equation}
		Now the reduced density matrix $\tilde{\rho}_A\equiv\Tr_{\bar{A}}\tilde{\rho}$ becomes
		\begin{equation}\label{RhoAtilde}
			\begin{aligned}
				\tilde{\rho}_A&=\Tr_{\bar{A}}\left[U_AU_{\bar{A}}\left(\rho_{A_1\bar{A}_1}\otimes \chi_{A_2\bar{A}_2}\right)U^\dagger_AU^\dagger_{\bar{A}}\right]\\
				&=\Tr_{\bar{A}}\left[U_A\left(\rho_{A_1\bar{A}_1}\otimes \chi_{A_2\bar{A}_2}\right)U^\dagger_A\right]\\
				&=U_A\Tr_{\bar{A}}\left(\rho_{A_1\bar{A}_1}\otimes \chi_{A_2\bar{A}_2}\right)U^\dagger_A\\
				&=U_A\left(\rho_{A_1}\otimes \chi_{A_2}\right)U^\dagger_A
			\end{aligned}
		\end{equation}
		where $\rho_{A_1}\equiv \Tr_{\bar{A}_1}\rho_{A_1\bar{A}_1}=\sum_{i}p_i\ket{i}_{A_1}\bra{i}_{A_1},~ p_i\equiv\sum_jp_{ij}$ and $\chi_{A_2}\equiv \Tr_{\bar{A}_2}\chi_{A_2\bar{A}_2}$. We then can calculate the corresponding entropy by using \eqref{RhoAtilde}:
		\begin{align}
			S(\tilde{\rho}_A)=S\left(U_A\left(\rho_{A_1}\otimes \chi_{A_2}\right)U^\dagger_A\right)=S\left(\rho_{A_1}\otimes \chi_{A_2}\right)=S(\rho_{A_1})+S(\chi_{A_2})
		\end{align}
		where the second equality makes use of the fact the von Neumann entropy is invariant under unitary transformations. Notice that $\tilde{\rho}_a\equiv\Tr_{\bar{a}}\tilde{\rho}=\sum_{i}p_i\ket{\tilde{i}}_{a}\bra{\tilde{i}}_{a}$ has exactly the same eigenvalues as $\rho_{A_1}=\sum_{i}p_i\ket{i}_{A_1}\bra{i}_{A_1}$ such that 
		\begin{equation}
			S(\rho_{A_1})=-\sum_{i}p_i\log p_i=S(\tilde{\rho}_a)
		\end{equation}
		Therefore, after identifying $S(\chi_{A_2})$ as $\mathcal{L}_{A}$ we obtain
		\begin{align}
			S(\tilde{\rho}_A)=\mathcal{L}_A+S(\tilde{\rho}_a)
		\end{align}
		and likewise for $\tilde{\rho}_{\bar{A}}\equiv\Tr_{A}\tilde{\rho}$ and $\tilde{\rho}_{\bar{a}}\equiv\Tr_{a}\tilde{\rho}$, we have
		\begin{equation}
			S(\tilde{\rho}_{\bar{A}})=\mathcal{L}_{\bar{A}}+S(\tilde{\rho}_{\bar{a}})
		\end{equation}
		where $\mathcal{L}_{\bar{A}}=S(\chi_{\bar{A}_2}),~\chi_{\bar{A}_2}\equiv \Tr_{A_2}\chi_{A_2\bar{A}_2}$. Notice that $\chi_{A_2\bar{A}_2}=\ket{\chi}_{A_2\bar{A}_2}\bra{\chi}_{A_2\bar{A}_2}$ is a pure state on $A_2\bar{A}_2$ by definition, then we know that $S(\chi_{\bar{A}_2})=S(\chi_{A_2})$, or equivalently,
		\begin{equation}
			\mathcal{L}_{A}=\mathcal{L}_{\bar{A}}
		\end{equation}

		\item $(2)\Rightarrow(3)$: This part follows the same idea in \cite{Jafferis:2015del,Dong:2016eik}.
		Before we start, we recall how von Neumann entropy varies when we perturb the density matrix:
		\begin{equation}
			\delta S(\rho)\equiv S(\rho+\delta \rho)-S(\rho)\approx-\Tr\left(\delta\rho\log\rho\right)
		\end{equation}
		Now we can proceed. Given a density matrix $\tilde{\sigma}$ on $\H_{code}$ and its reduced density matrices $\tilde{\sigma}_A, \tilde{\sigma}_a$, using \eqref{RT1} we have
		\begin{align}
			S(\tilde{\sigma}_A)=\Tr_a\left(\tilde{\sigma}_a\mathcal{L}_A\right)+S(\tilde{\sigma}_a)
		\end{align}
		where we use $\Tr_a\tilde{\sigma}_a=1$ and promote $\mathcal{L}_A$ to an operator of the form $\mathcal{L}_A I_{a}\in \mathcal{L}(\H_a)$. Now we expand in terms of $\tilde{\sigma}$ in first order:
		\begin{equation}
			\Tr_A \left( \delta \tilde{\sigma}_A \tilde{K}^\sigma_A\right)=\Tr_a\left[\delta \tilde{\sigma}_a\left(\tilde{K}^\sigma_a+\mathcal{L}_A I_{a}\right)\right]
		\end{equation}
		where $\tilde{K}^\sigma_A=-\log \tilde{\sigma}_{A}$ and $\tilde{K}^\sigma_a=-\log \tilde{\sigma}_{a}$ are modular Hamiltonians. Both sides are linear in $\delta\tilde{\sigma}$ such that we can integrate it to $\tilde{\rho}$:
		\begin{equation}
			\Tr_A \left(  \tilde{\rho}_A \tilde{K}^\sigma_A\right)=\Tr_a\left[ \tilde{\rho}_a\left(\tilde{K}^\sigma_a+\mathcal{L}_A I_{a}\right)\right]=\Tr_a \left(\tilde{\rho}_a\tilde{K}^\sigma_a\right)+\mathcal{L}_A 
		\end{equation}
		Now with \eqref{RT1} we find that
		\begin{align*}
			S(\tilde{\rho}_A|\tilde{\sigma}_A)
			&=\Tr_A \left(  \tilde{\rho}_A \tilde{K}^\sigma_A\right)-S(\tilde{\rho}_A)\\
			&=\Tr_a \left(\tilde{\rho}_a\tilde{K}^\sigma_a\right)+\mathcal{L}_A -\left(\mathcal{L}_A+S(\tilde{\rho}_a)\right)\\
			&=\Tr_a \left(\tilde{\rho}_a\tilde{K}^\sigma_a\right)-S(\tilde{\rho}_a)\\
			&=S(\tilde{\rho}_a|\tilde{\sigma}_a)
		\end{align*}
		Similarly, we also have
		\begin{align}
			S(\tilde{\rho}_{\bar{A}}|\tilde{\sigma}_{\bar{A}})&=S(\tilde{\rho}_{\bar{a}}|\tilde{\sigma}_{\bar{a}})
		\end{align}
		
		\item $(3)\Rightarrow (4)$: 
		The relative entropy vanishes if and only if the two density matrices coincide, so if provided \eqref{relSeq2}, we are given a condition that
		\begin{equation} \label{condrhosigma}
			\tilde{\rho}_{\bar{A}}=\tilde{\sigma}_{\bar{A}}\Leftrightarrow\tilde{\rho}_{\bar{a}}=\tilde{\sigma}_{\bar{a}}
		\end{equation}
		
		Let us consider a state $\ket{\tilde{\phi}}\in \H_{code}$ and a Hermitian operator $\tilde{O}_a\in\mathcal{L}(\H_a)$, we can construct a new state $\ket{\tilde{\psi}}$ as follows,
		\begin{equation}
			\ket{\tilde{\psi}}\equiv e^{i\tilde{O}_a s}\ket{\tilde{\phi}}
		\end{equation}
		for some real constant $s$. We now define two density matrices on $\H_{code}$:
		\begin{equation}
			\tilde{\sigma}\equiv\ket{\tilde{\psi}}\bra{\tilde{\psi}},\quad \tilde{\rho}\equiv\ket{\tilde{\phi}}\bra{\tilde{\phi}}
		\end{equation}
		then we have
		\begin{align*}
			\tilde{\sigma}_{\bar{a}}
			&=\Tr_a\ket{\tilde{\psi}}\bra{\tilde{\psi}}\\
			&=\Tr_a\left[e^{i\tilde{O}_a s}\ket{\tilde{\phi}}\bra{\tilde{\phi}}e^{-i\tilde{O}_a s}\right]\\
			&=\Tr_a\ket{\tilde{\phi}}\bra{\tilde{\phi}}\\
			&=\tilde{\rho}_{\bar{a}}
		\end{align*}
		By condition \eqref{condrhosigma}, we must have
		\begin{equation}
			\tilde{\rho}_{\bar{A}}=\tilde{\sigma}_{\bar{A}}
		\end{equation}
		which further implies
		\begin{align}
			&\Tr_{\bar{A}}\left(\tilde{\rho}_{\bar{A}}X_{\bar{A}}\right)=\Tr_{\bar{A}}\left(\tilde{\sigma}_{\bar{A}}X_{\bar{A}}\right),\quad \forall X_{\bar{A}}\in \mathcal{L}(\H_{\bar{A}})\\
			\Rightarrow \quad & \Tr_{A\bar{A}}\left(\ket{\tilde{\phi}}\bra{\tilde{\phi}}X_{\bar{A}}\right)=\Tr_{A\bar{A}}\left(\ket{\tilde{\psi}}\bra{\tilde{\psi}}X_{\bar{A}}\right)\\
			\Rightarrow \quad &\bra{\tilde{\phi}}X_{\bar{A}}\ket{\tilde{\phi}}=\bra{\tilde{\psi}}X_{\bar{A}}\ket{\tilde{\psi}}
		\end{align}
		For infinitesimal $\lambda$, we expand the right-hand side of the last result in first order to obtain
		\begin{equation}
			\bra{\tilde{\phi}}[\tilde{O}_a,X_{\bar{A}}]\ket{\tilde{\phi}}=0,\quad \forall X_{\bar{A}}\in \mathcal{L}(\H_{\bar{A}})
		\end{equation}
		Notice that the above derivation requires $\tilde{O}_a$ to be Hermitian, but any operator can be rewritten as a complex linear combination of Hermitian operators as follows,
		\begin{equation}\label{HermiticityExpansion}
			\tilde{O}_a=\frac{\tilde{O}_a+\tilde{O}^\dagger_a	}{2}+i\frac{\tilde{O}_a-\tilde{O}^\dagger_a	}{2i}
		\end{equation}
		where $\frac{\tilde{O}_a+\tilde{O}^\dagger_a	}{2}$ and $\frac{\tilde{O}_a-\tilde{O}^\dagger_a	}{2i}$ are both Hermitian. Therefore, it is suffice to derive our result by using Hermitian operators only.

		Similarly, if we start with \eqref{relSeq1}, we will obtain
		\begin{equation}
			\bra{\tilde{\phi}}[\tilde{O}_{\bar{a}},X_{A}]\ket{\tilde{\phi}}=0,\quad \forall X_{A}\in \mathcal{L}(\H_{A})
		\end{equation}
		for any operator $\tilde{O}_{\bar{a}}\in\mathcal{L}(\H_{\bar{a}})$.

		\item $(4)\Rightarrow(1)$: 
		Before our proof, we outline our strategy. We first introduce two auxiliary systems $R,\bar{R}$ with $\abs{R}=\abs{a},\abs{\bar{R}}=\abs{\bar{a}}$ and a state $\ket{\phi}\in \H_R\otimes\H_{\bar{R}}\otimes\H_{code}\subseteq \H_R\otimes\H_{\bar{R}}\otimes\H$ defined by
		\begin{equation}
			\ket{\phi}\equiv \frac{1}{\sqrt{\abs{R}\abs{\bar{R}}}}\sum_{i,j}\ket{i}_R\ket{j}_{\bar{R}}\ket{\tilde{ij}}
		\end{equation}
		Next we introduce a set of orthonormal states $\ket{\chi_j}_{A_2\bar{A}}\in\H_{A_2\bar{A}}$ and $\ket{\chi}_{\bar{R}A_2\bar{A}}\equiv \frac{1}{\sqrt{\abs{\bar{R}}}}\sum_{j}\ket{j}_{\bar{R}}\ket{\chi_j}_{A_2\bar{A}}$, then we define a state $\ket{\phi^\p}\in\H_R\otimes\H_{\bar{R}}\otimes\H$ by
		\begin{equation}
			\ket{\phi^\p}\equiv \frac{1}{\sqrt{\abs{R}\abs{\bar{R}}}}\sum_{i,j}\ket{i}_R\ket{j}_{\bar{R}}\ket{i}_{A_1}\ket{\chi_j}_{A_2\bar{A}}=\frac{1}{\sqrt{\abs{R}}}\sum_{i}\ket{i}_R\ket{i}_{A_1}\ket{\chi}_{\bar{R}A_2\bar{A}}
		\end{equation}
		These two states corresponds to two pure density matrices $\rho\equiv \ket{\phi}\bra{\phi},~\rho^\p\equiv \ket{\phi^\p}\bra{\phi^\p}$. To prove that there exists a unitary operator $U_A\in \mathcal{L}(\H_{A})$ such that 
		\begin{equation}
			\ket{\tilde{ij}}=U_A\left(\ket{i}_{A_1}\ket{\chi_j}_{A_2\bar{A}}\right)
		\end{equation}
		we can instead define $\rho_{\bar{R}R\bar{A}}=\Tr_A\rho$ then try to prove that
		\begin{equation}\label{condPurification}
			\Tr_A\rho^\p=\rho_{\bar{R}R\bar{A}}
		\end{equation}
		which tells us that $\rho^\p,\rho$ are both purifications of $\rho_{\bar{R}R\bar{A}}$. Two different purifications differ by a unitary transformation, i.e. there exist a unitary $U_A\in\L(\H_A)$ such that
		\begin{equation}
			\ket{\phi}=U_A\ket{\phi^\p}\quad \Rightarrow \quad \ket{\tilde{ij}}=U_A\left(\ket{i}_{A_1}\ket{\chi_j}_{A_2\bar{A}}\right)
		\end{equation} 
		which will end the proof.

		We digress a bit to mention a useful statement. It is possible to find an alternative operator which acts on different space but still does the same work as the original one. To be more specific, consider a bipartite system $XY$ with an arbitrary state $\ket{\psi}\in\H_{X}\otimes\H_{Y}$ which is in general of the form
		\begin{equation}
			\ket{\psi}=C^{ab}\ket{a}_X\ket{b}_Y
		\end{equation}
		where $\ket{a}_X, \ket{b}_Y$ represents bases in respective spaces and Einstein summation is understood. Suppose we have operators $O_X\in\H_X, O_Y\in\H_Y$ which satisfy
		\begin{equation}
			O_X\ket{\psi}=O_Y\ket{\psi}
		\end{equation}
		i.e. they are identical when acting on states in $\ket{\psi}\in\H_{X}\otimes\H_{Y}$. We would like to see how the components of the two operators are related. We further assume that
		\begin{equation}
			O_X\ket{a}_X=O^{ca}_X\ket{c}_X,\quad O_Y\ket{b}_Y=O^{db}_Y\ket{d}_X
		\end{equation}
		where $O^{ca}_X,O^{db}_Y$ are constants. We then have
		\begin{align*}
			O_X\ket{\psi}
			&=C^{ab}O^{ca}_X\ket{c}_X\ket{b}_Y\\
			=O_Y\ket{\psi}&=C^{ab}O_Y^{db}\ket{a}_X\ket{d}_Y=C^{ca}O_Y^{ba}\ket{c}_X\ket{b}_Y
		\end{align*}
		After modifying dummy indices, we obtain
		\begin{equation}
			O^{ca}_XC^{ab}=C^{ca}O_Y^{ba}
		\end{equation}
		or in matrix notation
		\begin{equation}
			O_XC=CO_Y^T
		\end{equation}
		This result tells us that how we can find a new operator which acts on the complementary space but does the same work as the given one.

		Now back to our business. Given \eqref{ComO1} for an arbitrary operator $\tilde{O}_a\in\L(\H_a)$, we have:
		\begin{equation}
			\bra{\tilde{ij}}[\tilde{O}_a,X_{\bar{A}}]\ket{\tilde{ij}}=0,\quad \forall X_{\bar{A}}\in \mathcal{L}(\H_{\bar{A}})
		\end{equation}
		By the statement we just mentioned, we can find an alternative operator $O_R$ which satisfies
		\begin{equation}
			O_R\ket{\phi}=\tilde{O}_a\ket{\phi}
		\end{equation} 
		such that
		\begin{equation}
			\bra{\phi}[O_R,X_{\bar{A}}]\ket{\phi}=0
		\end{equation}
		We can further introduce an operator acting on different space in the above equality:
		\begin{equation}
			\bra{\phi}[O_R,X_{\bar{A}}Y_{\bar{R}}]\ket{\phi}=0,\quad \forall Y_{\bar{R}}\in\H_{\bar{R}}
		\end{equation}
		because $Y_{\bar{R}}$ commutes with both $O_R$ and $X_{\bar{A}}$. It implies that
		\begin{align*}
			\Tr_{\bar{R}R\bar{A}}\left(\rho_{\bar{R}R\bar{A}}[O_R,X_{\bar{A}}Y_{\bar{R}}]\right)
			&=\Tr\left(\rho[O_R,X_{\bar{A}}Y_{\bar{R}}]\right)\\
			&=\bra{\phi}[O_R,X_{\bar{A}}Y_{\bar{R}}]\ket{\phi}\\
			&=0
		\end{align*}
		where we define $\rho=\ket{\phi}\bra{\phi}$ and $\rho_{\bar{R}R\bar{A}}=\Tr_A\rho$. Rearranging left-hand side gives
		\begin{align*}
			\Tr_{\bar{R}R\bar{A}}\left(\rho_{\bar{R}R\bar{A}}O_RX_{\bar{A}}Y_{\bar{R}}\right)&=\Tr_{\bar{R}R\bar{A}}\left(\rho_{\bar{R}R\bar{A}}X_{\bar{A}}Y_{\bar{R}}O_R\right)\\
			&=\Tr_{\bar{R}R\bar{A}}\left(O_R\rho_{\bar{R}R\bar{A}}X_{\bar{A}}Y_{\bar{R}}\right)
		\end{align*}
		which further implies
		\begin{equation}
			[\rho_{\bar{R}R\bar{A}},O_R\otimes I_{\bar{R}\bar{A}}]=0, \quad \forall\tilde{O}_a\in\L(\H_a)
		\end{equation}
		It constraints $\rho_{\bar{R}R\bar{A}}$ to be of the form
		\begin{equation}
			\rho_{\bar{R}R\bar{A}}=I_R\otimes M_{\bar{R}\bar{A}}
		\end{equation}
		where $M_{\bar{R}\bar{A}}$ is an operator to be determined. We note that
		\begin{equation}
			\rho_{\bar{R}\bar{A}}=\Tr_R\rho_{\bar{R}R\bar{A}}=\abs{R}M_{\bar{R}\bar{A}}\quad\Rightarrow\quad  M_{\bar{R}\bar{A}}=\frac{1}{\abs{R}}\rho_{\bar{R}\bar{A}}
		\end{equation}
		such that
		\begin{equation}\label{condPurification2}
			\rho_{\bar{R}R\bar{A}}=\rho_R\otimes \rho_{\bar{R}\bar{A}},\quad  \rho_R\equiv\frac{1}{\abs{R}}I_R
		\end{equation}
		As for $\rho^\p$, we can directly compute its trace over $A$:
		\begin{align}
			&\rho^\p=\frac{1}{\abs{R}}\sum_{i,j}\left(\ket{i}_R\bra{j}_R\right)\otimes \left(\ket{i}_{A_1}\bra{j}_{A_1}\right)\otimes \chi\\
			\Rightarrow\quad&\Tr_A\rho^\p=\rho_R\otimes \Tr_{A_2}\chi
		\end{align}
		where $\chi\equiv\ket{\chi}_{\bar{R}A_2\bar{A}}\bra{\chi}_{\bar{R}A_2\bar{A}}$. If we compare it with \eqref{condPurification2}, we find that condition \eqref{condPurification} now reduces to
		\begin{equation}
			\Tr_{A_2}\chi=\rho_{\bar{R}\bar{A}}
		\end{equation}
		i.e. $\chi$ is a purification of $\rho_{\bar{R}\bar{A}}$ by introducing $A_2$. Now our goal becomes proving that purifying $\rho_{\bar{R}\bar{A}}$ by $A_2$ is possible, which is true as long as $\abs{A_2}\geq \rank(\rho_{\bar{R}\bar{A}})$.

		Notice that \eqref{condPurification2} tells us that 
		\begin{equation}
			\rank(\rho_{\bar{R}R\bar{A}})=\rank(\rho_R)\rank(\rho_{\bar{R}\bar{A}})=\abs{R}\rank(\rho_{\bar{R}\bar{A}})
		\end{equation}
		
		Since $\rho$ can be regarded as a purification of $\rho_{\bar{R}R\bar{A}}$, we must have
		\begin{equation}
			\abs{A}\geq \rank(\rho_{\bar{R}R\bar{A}})=\abs{R}\rank(\rho_{\bar{R}\bar{A}})
		\end{equation}
		Also by assumption $\abs{A}=\abs{A_1}\abs{A_2}$ and $\abs{R}=\abs{a}=\abs{A_1}$, we further have
		\begin{equation}
			\abs{A_1}\abs{A_2}\geq \abs{A_1}\rank(\rho_{\bar{R}\bar{A}})\Rightarrow\abs{A_2}\geq \rank(\rho_{\bar{R}\bar{A}})
		\end{equation} 
		Therefore, it is possible to purify $\rho_{\bar{R}\bar{A}}$ by subsystem $A_2$, which concludes our proof, and we can prove \eqref{ijtilde2} in exactly the same way.

		\item $(1)\Rightarrow(5)$: Here we give a constructive proof. For any operator $\tilde{O}_a\in\mathcal{L}(\H_a)$, it is defined by what it can do to the basis, i.e. 
		\begin{equation}
			\tilde{O}_a\ket{\tilde{i}}_a\equiv O^{ji}_{a}\ket{\tilde{j}}_a
		\end{equation}
		where $O^{ji}_{a}$ are some constants.
		We then introduce another operator $O_{A_1}\in\L(\H_{A_1})$ by using the same coefficients:
		\begin{equation}
			O_{A_1}\ket{i}_{A_1}\equiv O^{ji}_{a}\ket{j}_{A_1}
		\end{equation}
		which can always be done because we assume $\abs{A_1}=\abs{a}$. Finally, given \eqref{ijtilde1} we define
		\begin{equation}
			O_A\equiv U_A O_{A_1}U_A^\dagger
		\end{equation}
		which satisfies:
		\begin{align*}
			O_A\ket{\tilde{ij}}
			&= U_A O_{A_1}U_A^\dagger U_A\left(\ket{i}_{A_1}\ket{\chi_j}_{A_2\bar{A}}\right)\\
			&= U_A O_{A_1}\left(\ket{i}_{A_1}\ket{\chi_j}_{A_2\bar{A}}\right)\\
			&=  O^{ki}_{a}U_A\left(\ket{k}_{A_1}\ket{\chi_j}_{A_2\bar{A}}\right)\\
			&=  O^{ki}_{a}\ket{\tilde{kj}}\\
			&=  \tilde{O}_{a}\ket{\tilde{ij}}
		\end{align*}
		which further implies \eqref{correct1}, and the same procedure works for \eqref{correct2}.

		\item $(5)\Rightarrow(4)$: The proof is straightforward. Given \eqref{correct1}, we have
		\begin{equation}
			\bra{\tilde{\phi}}[\tilde{O}_a,X_{\bar{A}}]\ket{\tilde{\phi}}=\bra{\tilde{\phi}}[O_A,X_{\bar{A}}]\ket{\tilde{\phi}}=0,\quad \forall X_{\bar{A}}\in\mathcal{L}(\H_{\bar{A}})
		\end{equation}
		where the last equality holds because $O_A$ and $X_{\bar{A}}$ acts on different spaces. The same procedure works for \eqref{ComO2}.

	\end{itemize}

\end{proof}

\bibliographystyle{JHEP}
\bibliography{bib}
\end{document}